%% file: 2dlj-7.tex
\documentclass[aps,prb,superscriptaddress,reprint]{revtex4-1}
\usepackage[permil]{overpic}
\usepackage{graphicx}
\usepackage{latexsym}
\usepackage{amsmath}
\usepackage{amssymb}
\usepackage{upgreek}
\usepackage{xcolor}
\usepackage[colorlinks]{hyperref}
\hypersetup{allcolors=blue}
\begin{document}
\title{The precision of molecular dynamics simulations and what we can learn from it?}
\author{M. V. Kondrin}
\email{mkondrin@hppi.troitsk.ru}
\affiliation{Institute for High Pressure Physics RAS, 108840 Troitsk, Moscow, Russia}
\author{Y.B. Lebed}
\affiliation{Institute for Nuclear Research RAS, 117312 Moscow, Russia}
\begin{abstract}
We have investigated  by molecular dynamics method  the influence of a finite number of particles used in computer simulations on  fluctuations of thermodynamic properties. As a case study, we  used the two-dimensional Lennard-Jones system.  2D Lennard-Jones system, besides being an archetypal one,  is a subject of long debate, as to whether it has continuous (infinite-order) or discontinuous (first-order) melting transition. We have found  that  anomalies on the equation of state (the van-der-Waals or Myer-Wood loops)  previously considered a hallmark of the first order phase transition, are at best at the level of noise, since they have the same magnitude as the  amplitude of pressure fluctuations. So, they could be regarded as statistically unsignificant effect. Also, we estimated  inherent statistical noise, present in computer simulations, and  came to conclusion, that it is larger, than predicted by statistical physics, and the difference between them (called algorithmic fluctuations) is possibly due to the computer-related issues. It was demonstrated that these fluctuations in principle could be observed in real-life physical experiments which would lead to practical resolution of The Matrix hypothesis.
\end{abstract}
\maketitle
\section{Introduction}
Fluctuations are an  inevitable feature of finite size systems' thermal behavior, which manifested itself in the Brownian motion of small particles. In thermodynamic limit (infinite size of the system), fluctuations decrease, so we can safely deal with average thermodynamic parameters.

However, molecular dynamics and, in  general, any computer simulations, use quite small  finite size systems. Inevitable question arises with this, how close we are to the  thermodynamic limit . To answer  this question we have to estimate the system's fluctuations.
 
The problem  was previously acknowledged in Hickman and Mishin paper\cite{hikman:prb16}, but they considered  the temperature fluctuations only. Their description starts from outline of fundamental problem with temperature fluctuations in computer simulations,  which has no unanimous  explanation. The explanation can be found in well-known course of theoretical physics \cite{landafshitz:v} which gives a result that relative fluctuations of thermodynamic parameter are roughly inversely proportional to the square root of the number of system's particles. 

If we limit ourselves to description of  NVE ensemble (with constant number of particles N, volume V and energy E ), the  pressure and temperature of the system are expected to fluctuate around some average values. Classical statistical mechanics predicts \cite{landafshitz:v} that fluctuations of pressure P and temperature T of large enough system are distributed according to the Gauss law with standard deviations:

\begin{equation}
\Delta T^2=\frac{k_B T^2}{N c_v}
\label{eq:1} 
\end{equation}
\begin{equation}
\Delta P^2=-k_B T \left( \frac{\partial P}{\partial V} \right)_S=\frac{k_BT}{V}B_S
\label{eq:2} 
\end{equation}

Here, $k_B$ -- the Boltzmann constant, $c_V$ -- specific heat at constant volume  per particle, $B_S$ -- adiabatic bulk modulus, index S means constant entropy. The second formula in the case of ideal gas gives:

\begin{equation*}
\left(\frac{\Delta P}{P}\right)^2=\frac{\gamma}{N}
\end{equation*}

where $\gamma$ is an adiabatic exponent. This suggests the general law that relative fluctuations of thermodynamic parameters approximately scale  to the square root of the system size. This conclusion is corroborated  by computer simulations of  pressure fluctuations in non-ideal plasma\cite{saitov:ht14}. So for typical system size used in molecular dynamics simulations ($N \le 10^5$), relative fluctuations are slightly less than 1 \%. The situation gets worse for {\em ab-initio} molecular dynamic simulations, where number of atoms doesn't exceed several hundreds, where fluctuations can reach of about 10 \%.

\begin{figure}
\includegraphics[width=\columnwidth]{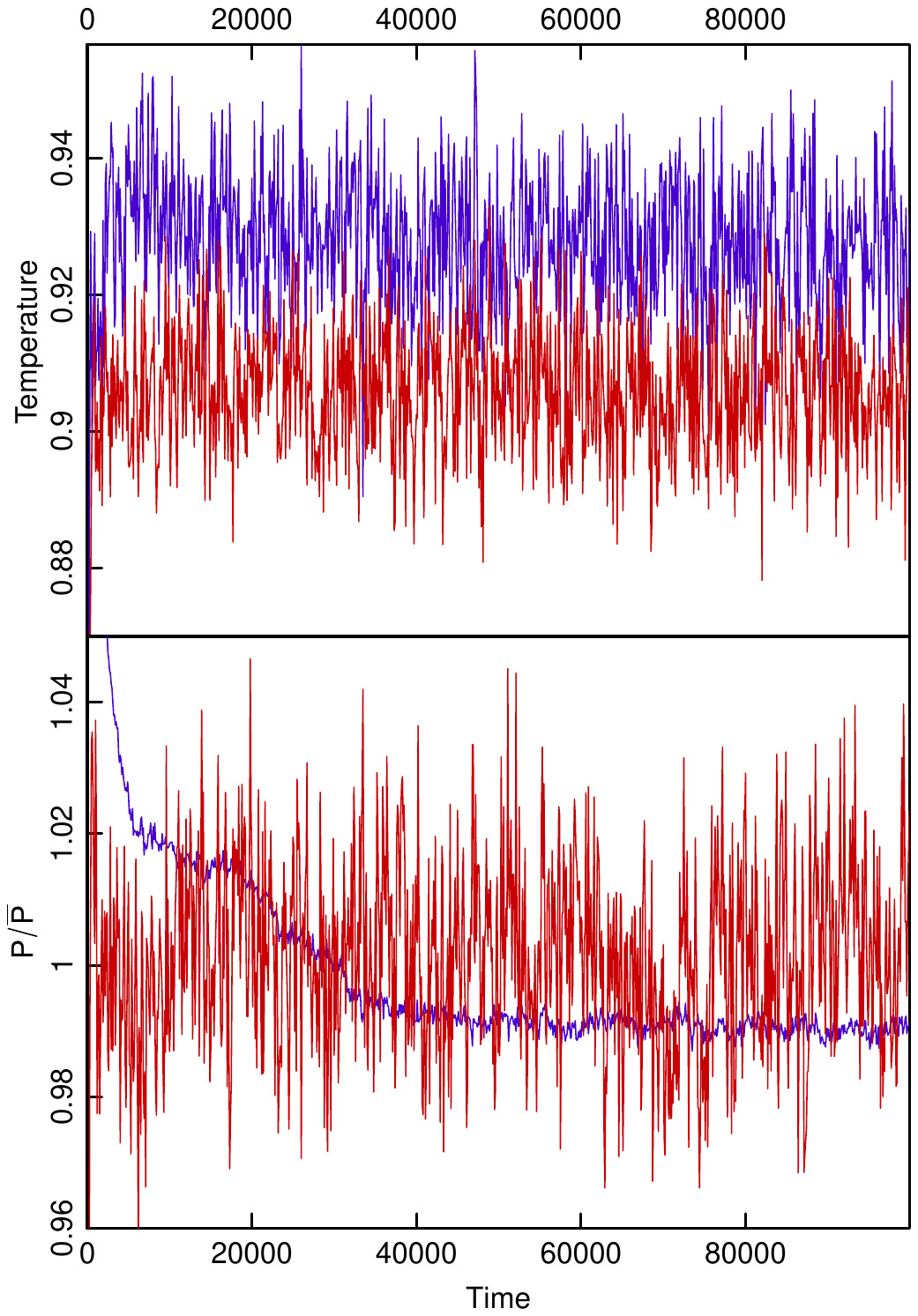}
\caption{Fluctuations of temperature (upper panel) and relative pressure (lower panel) at $T=0.9$ and two different densities $\rho=1.23$ (blue curves) and $\rho=0.86$ (red curves) which correspond to the states deep inside the crystal phase and in the liquid phase just below the transition respectively.}
\label{f:0}
\end{figure}

\begin{figure}
\includegraphics[width=\columnwidth]{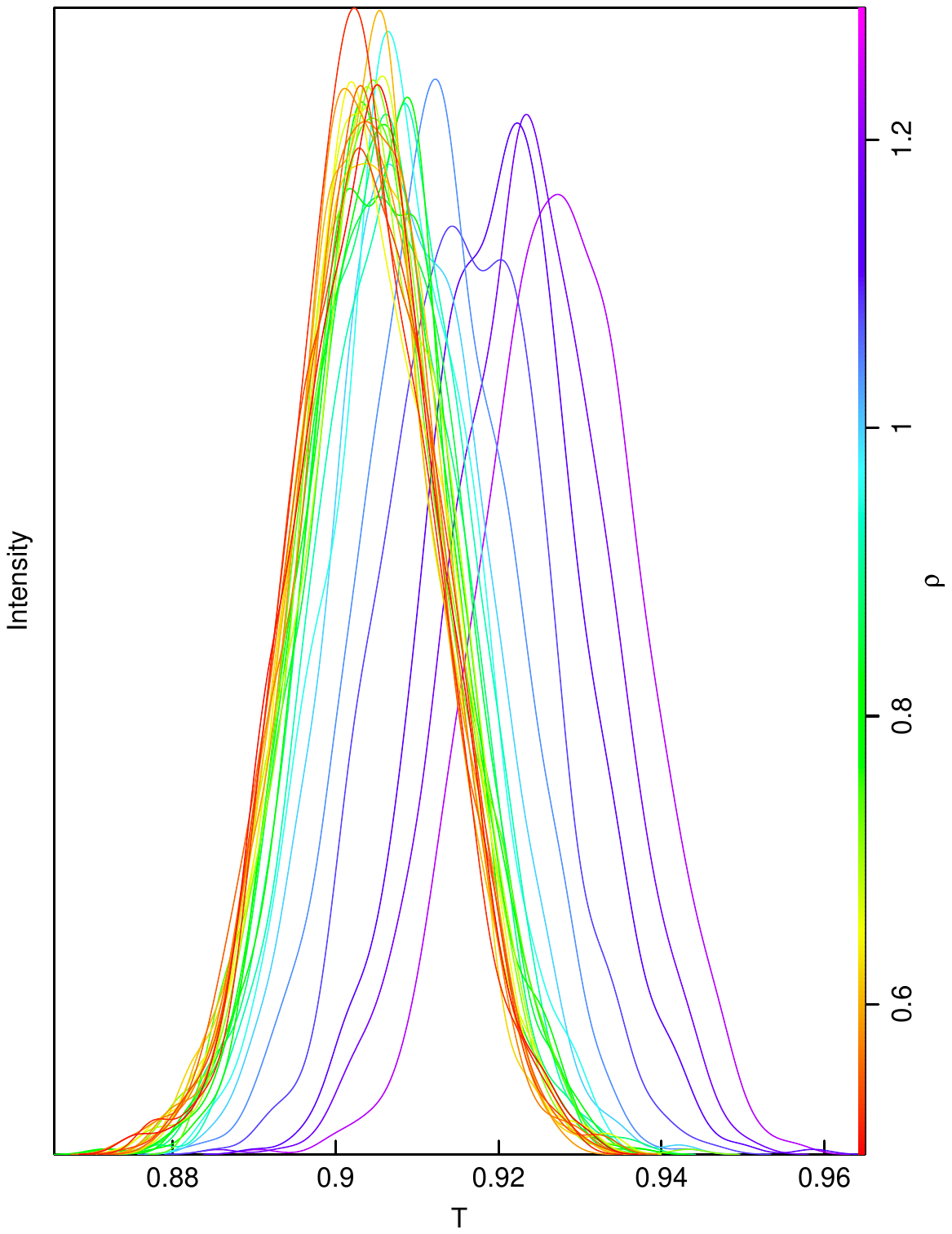}
\caption{Distribution density of thermal fluctuations at $T=0.9$ and different densities  with $N=10^4$ particles calculated in NVE ensemble. Density values are encoded in the color key at the right side of the plot}
\label{f:1}
\end{figure}

\begin{figure}
\includegraphics[width=\columnwidth]{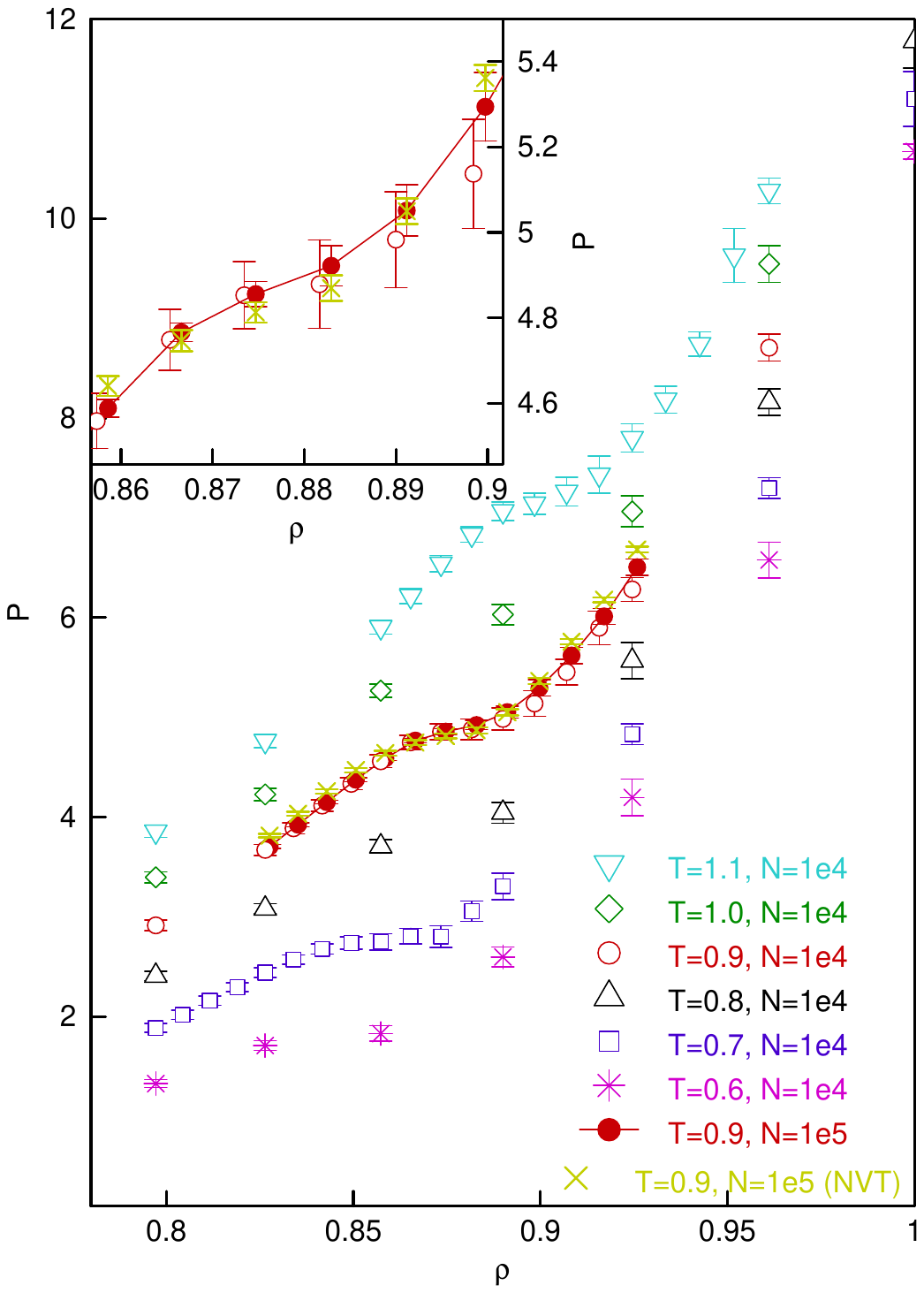}
\caption{Equation of state at different temperatures in the vicinity of melting transition. The inset shows enlarged region in the vicinity of transition at $T=0.9$ calculated for different ensembles and numbers of particles $N$ (see list in the main panel). The type of ensemble if not indicated in the plot means NVE ensemble.}
\label{f:2}
\end{figure}

To test these assumptions we have chosen  the two dimensional Lennard-Jones (2DLJ) system \cite{lj:wiki}. Being an archetypical model in computer simulations, its melting transition type attracts much attention .  Original  theory of 2D crystal  melting dating back to the 70s (the Berezinskii-Kosterlitz-Thouless-Halperin-Nelson-Young theory) suggests  that melting is an infinite-order, two-stage transition between crystal and liquid. The first stage -- crystal to  hexatic phase transition with the loss of translational order, and the second stage -- hexatic to liquid, with the loss of orientation order, see e.g. \cite{ryzhov:jetp23}. For this theory Kosterlitz and Thouless in 2016 were awarded with Nobel prize. For simplicity, we will  call it a continuous model. However, almost immediately this model was challenged by computer simulation study \cite{barker:pa81}  suggesting the first order phase transition in 2DLJ system. Subsequently, this system was addressed in the number of works \cite{frenkel:prl79,toxvaerd:pra81,tobochnik:prb82,koch:prb83,bakker:prl84,udink:prb87,chen:prl95,somer:prl97,somer:pre98} which produced controversial opinions ranging from first-order to  continuous character of melting transition. Hot debates followed through 1980-1990s  but did not lead to a clear conclusion. Echoes of the debates continue  until recently\cite{wierschem:pp10,patash:jpcc10,wierschem:prb11,haji:pre19,li:prl20,tsiok:jcp22}. Still, it is worth mentioning that  first order was predicted by the authors who used rather a small system $N < 10^4$ particles,  continuous transition was claimed by the authors who used a larger system up to $ N = 10^5$ and intermediate position - for the system with $N=2500$ particles \cite{patash:jpcc10}. The most recent paper using $N=512^2$ particles \cite{tsiok:jcp22}  favored the first-order transition. It seems that the authors' opinions are unconsciously effected by the scale of fluctuations . Most important are pressure fluctuations which influence the shape of equation of state (P-V) curve. Usually presence of van-der-Waals loops on the P-V curve (or the Mayer-Wood loop on  $P-\rho$  where $\rho$ is numerical density ) is  considered a decisive indicator of the first order transition. Still, the depth of the loops reported in many of these  works, is rather shallow and does not allow to  choose between the first-order or continuous phase transition . The most convincing would be comparison of the effect observed with the inherent system noise due to thermal fluctuations.

\section{Methods}
We tested validity of theoretical predictions by computations with classical molecular dynamics simulations of 2DLJ system as implemented in state of art LAMMPS software package \cite{plimpton:jcp95,lammps}. Input files from Carsten Svaneborg group home page \cite{cs:homepage} were  adapted to our needs. We used NVE ensemble with Langevin thermal dynamics with time step 0.01 LJ units. We used $5 \cdot 10^4$ steps for thermal equilibration and next $5 \cdot 10^4$ steps for estimation of thermodynamic parameters and their fluctuations . Validity of this choice will be explained later. As a rule we used $N=10^4$ particles. Some of the results were checked with larger  system containing $N=10^5$ particles. Also longer simulation times and NVT ensemble were used. Changing the ensemble to NVT with Nose-Hoover thermostat slightly influenced the result making the system fluctuate around  initially set temperature, but the amplitude of fluctuations  did not change. Starting configuration of particles was chosen random with  initial minimization of energy to avoid  overlapping particles. Random configuration taken by us illustrates crystallization of the liquid rather than crystal melting. The order of crystal-liquid phase transition  should not be influenced by the direction of the phase transition. The temperature was varied   in the range $T=0.6 -1.1$ with step 0.1 and density range $\rho=0.55-1.25$. To compare our findings with  previous results the temperature $T=3.0$ near the melting transition was investigated . The distribution density of thermal fluctuations was estimated by \verb|density| function with default parameters as implemented in R software package \cite{R}. 

\section{Results}

\begin{figure}
\includegraphics[width=\columnwidth]{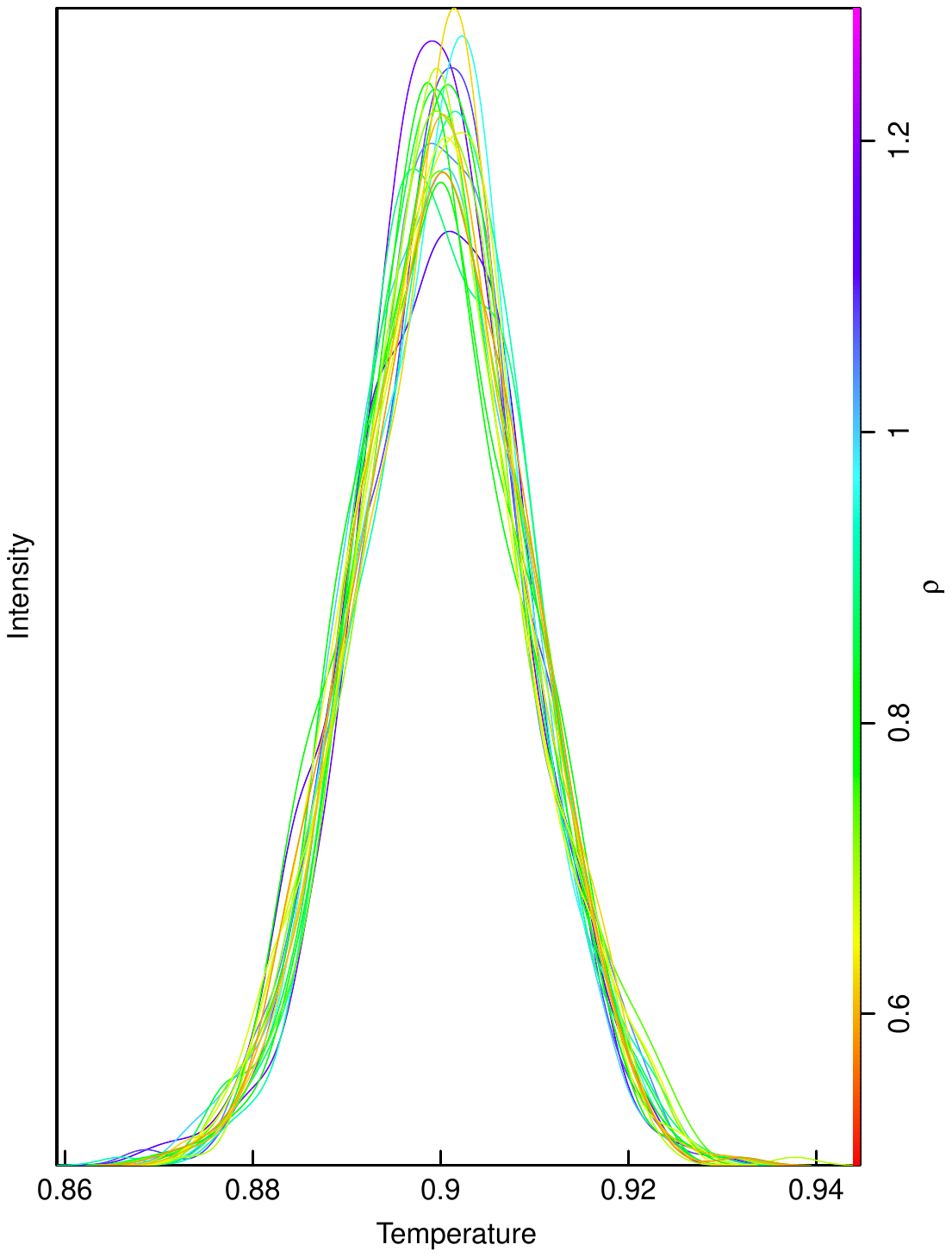}
\caption{Distribution density of thermal fluctuations at $T=0.9$ and different densities with $N=10^4$ particles calculated in NVT ensemble. Density values encoded in the color key at the right side of the plot }
\label{f:3}
\end{figure}

\begin{figure*}
\includegraphics[width=\textwidth]{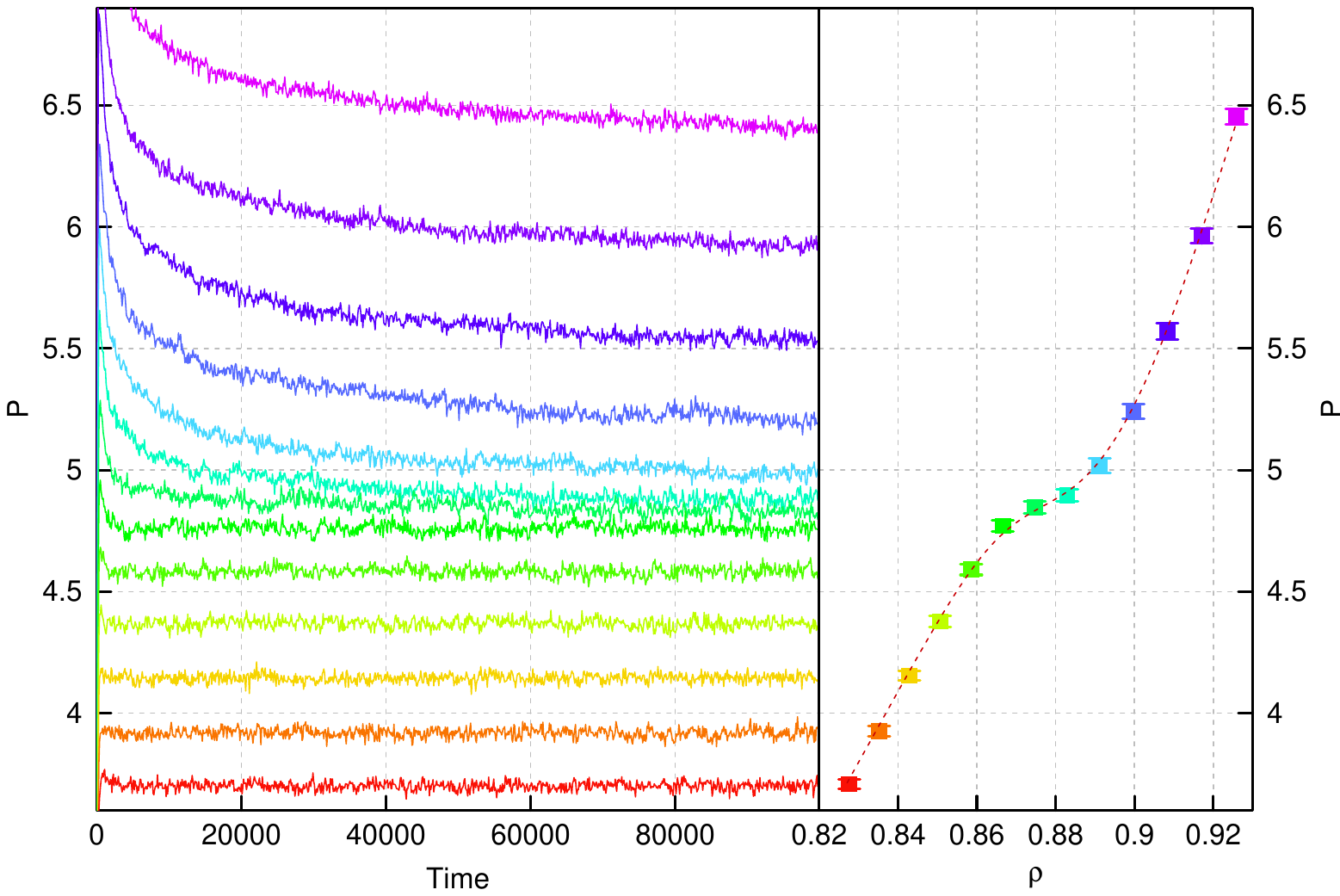}
\caption{Fluctuations of pressure in vicinity of melting transition at $T=0.9$ (left panel) and corresponding equation of state (right panel) for larger ensemble with $N=10^5$ particles.}
\label{f:4}
\end{figure*}

\begin{figure}
\includegraphics[width=\columnwidth]{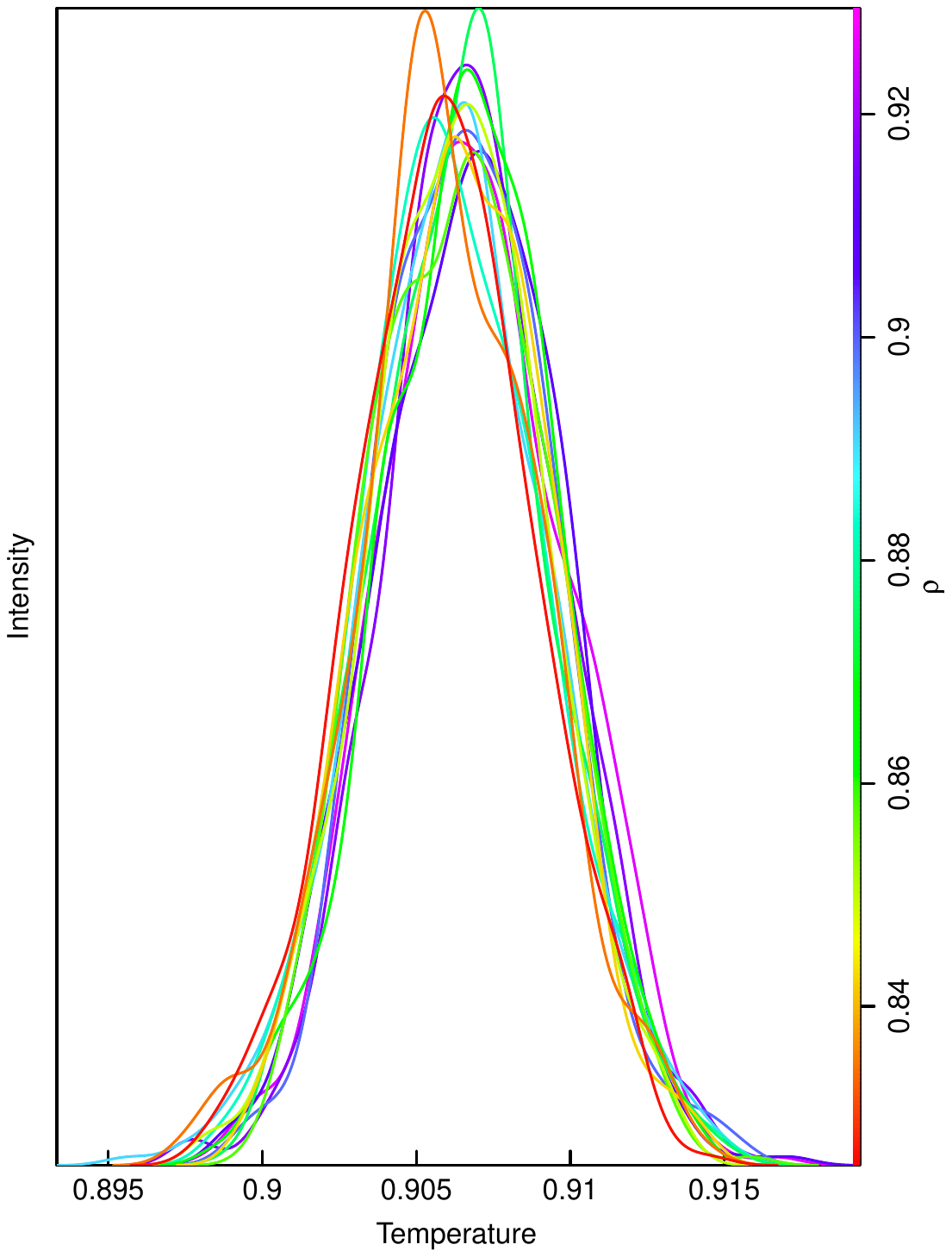}
\caption{Distribution density of thermal fluctuations in vicinity of melting transition at $T=0.9$ and different densities for larger ensemble with $N=10^5$ particles. The densities values are encoded in the color key at the right side of the plot.}
\label{f:5}
\end{figure}

\begin{figure}
\includegraphics[width=\columnwidth]{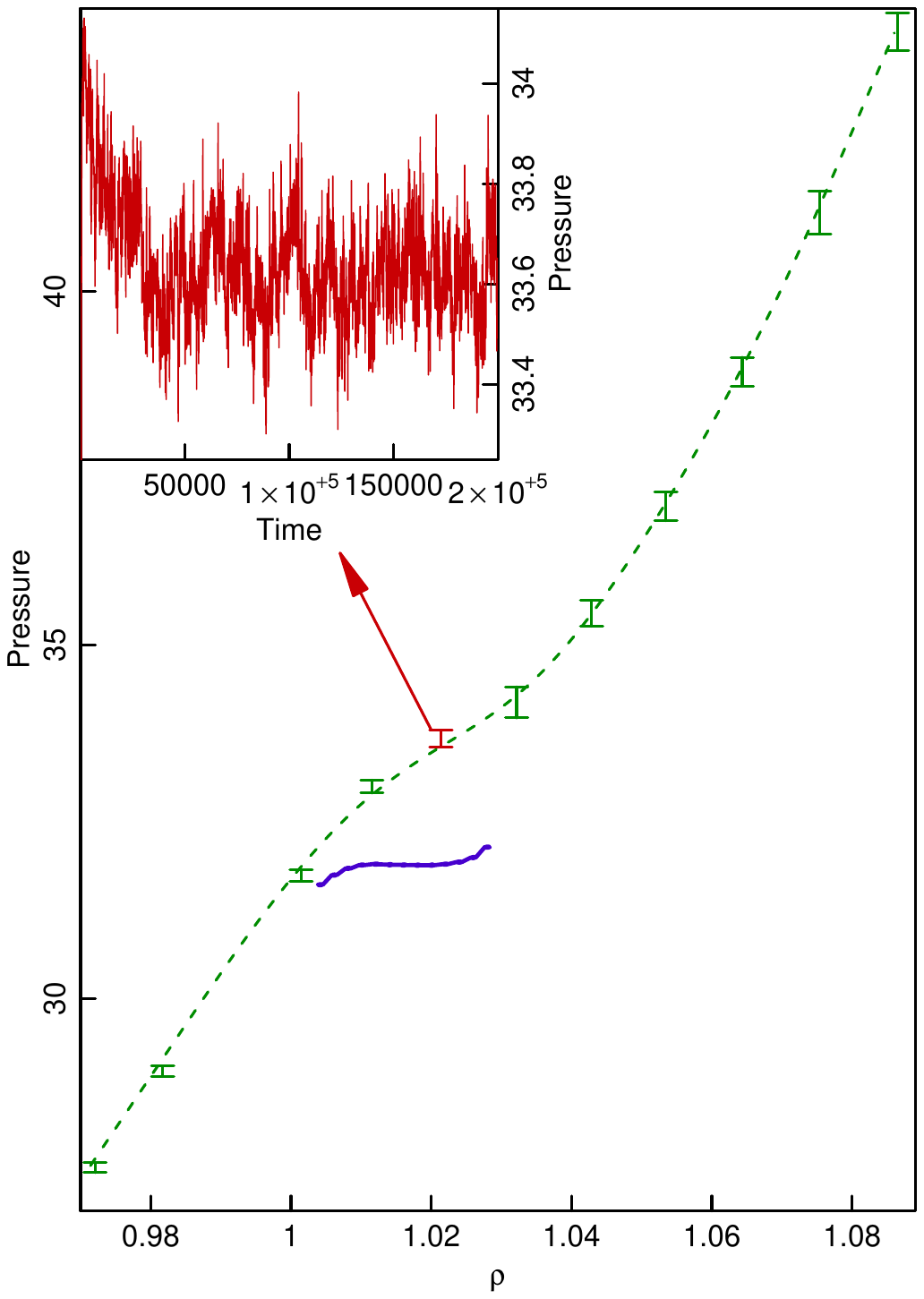}
\caption{Equation of state in vicinity of melting temperature calculated in the current work for NVE ensemble with $N=10^5$ particles at $T=3.0$ (green dashed curve is guide to the eyes). Blue curve is the data for NVT  ensemble for $N=256^2$ particles at the same temperature obtained in Ref.~\cite{tsiok:jcp22}. Inset shows temperature fluctuations for the chosen (marked in red) point in the middle of the melting transition.}
\label{f:6}
\end{figure}

\begin{figure*}
\includegraphics[width=\textwidth]{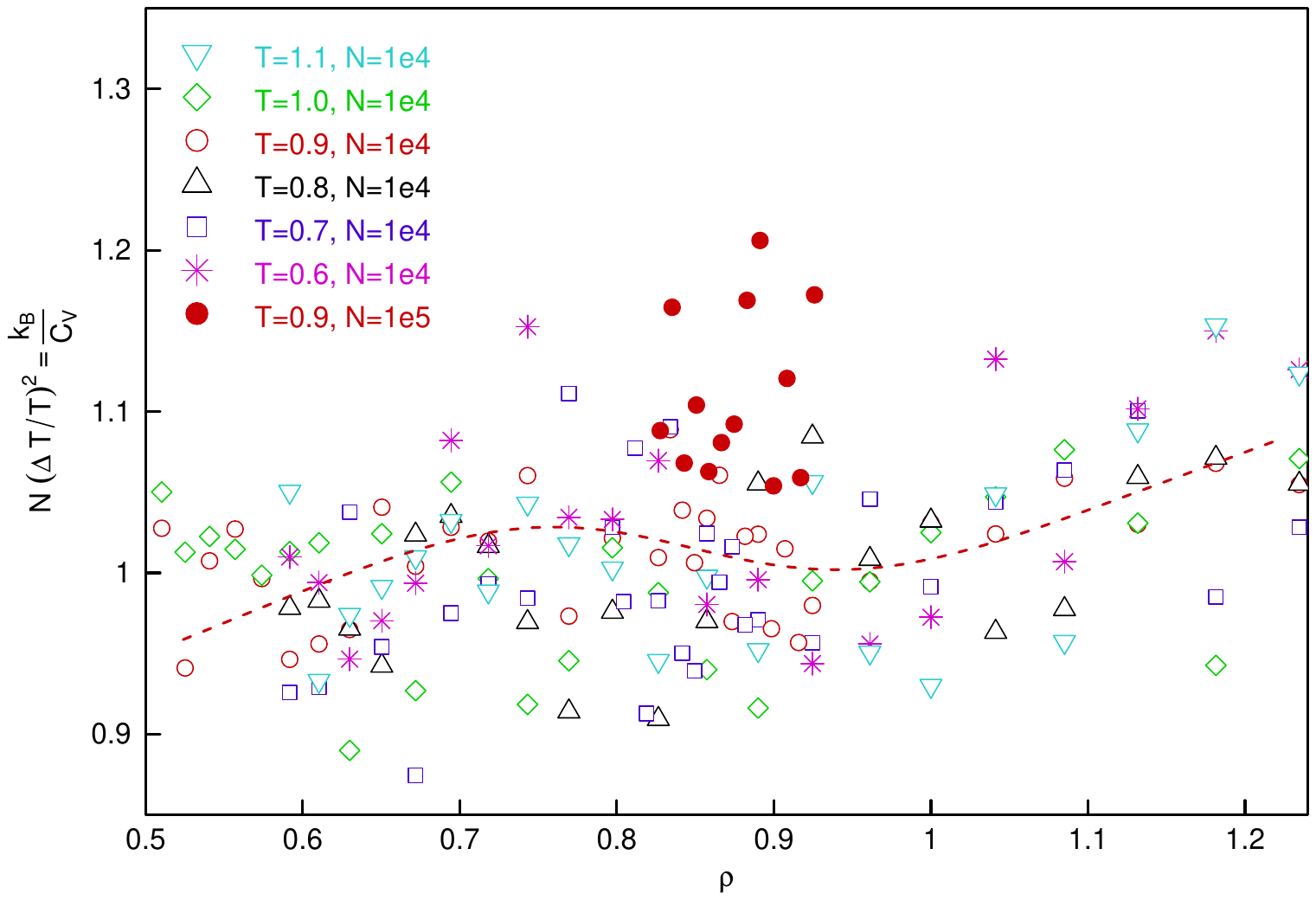}
\caption{Heat capacity of the 2DLJ system estimated from thermal fluctuations by Eq.~\ref{eq:1}. Dashed curve is guide to the eyes.}
\label{f:7}
\end{figure*}

\begin{figure}
\includegraphics[width=\columnwidth]{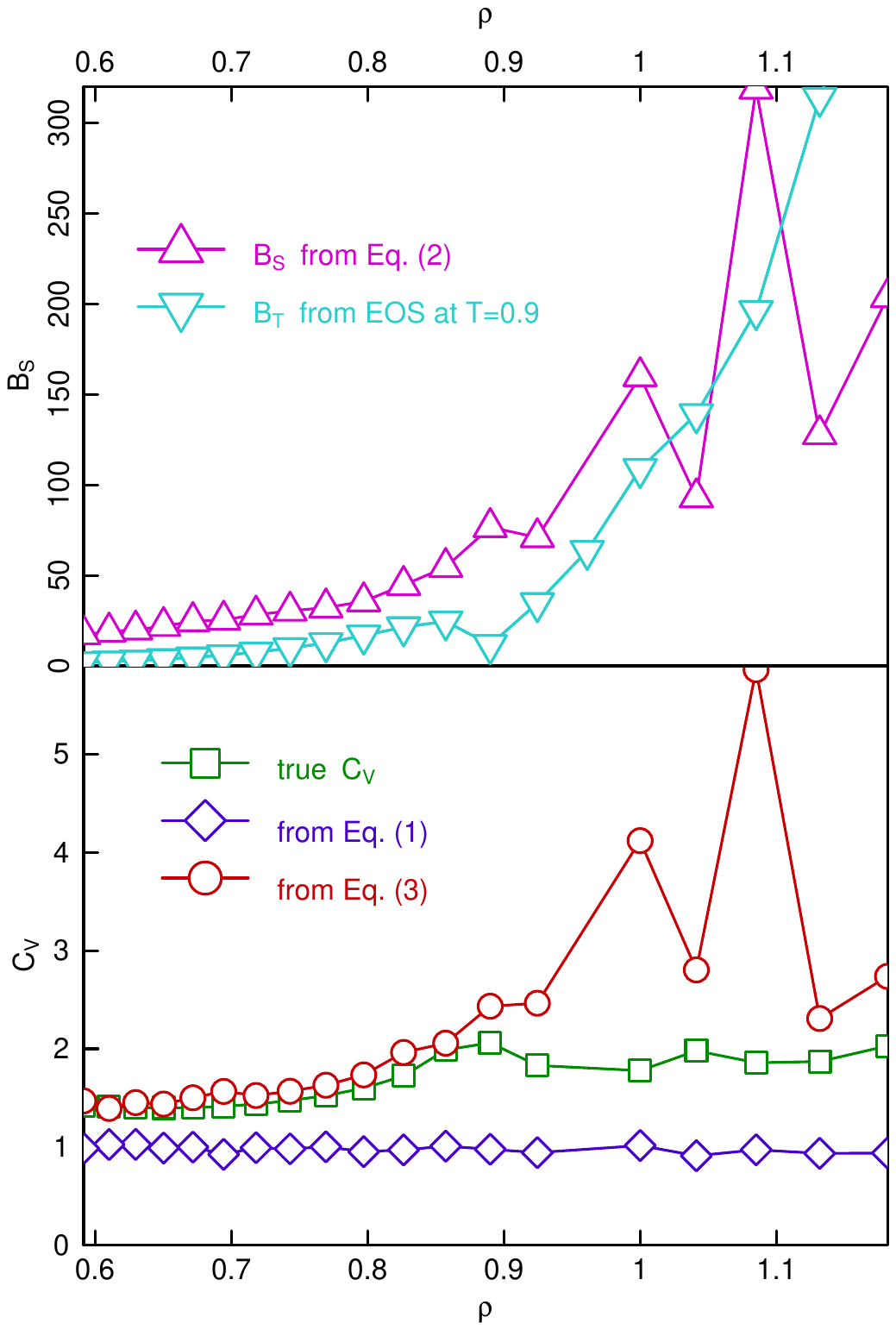}
\caption{Adiabatic bulk modulus $B_S$ and heat capacity $C_V$ evaluated from Eqs.~\ref{eq:1}-\ref{eq:3}. ``True'' $C_V$ and isothermal bulk modulus $B_T$ are evaluated by numerical differentiation of internal energy along isochores and equation of state $P(\rho)$ along isotherm $T=0.9$ respectively.}
\label{f:8}
\end{figure}

\begin{figure*}
\includegraphics[width=\textwidth]{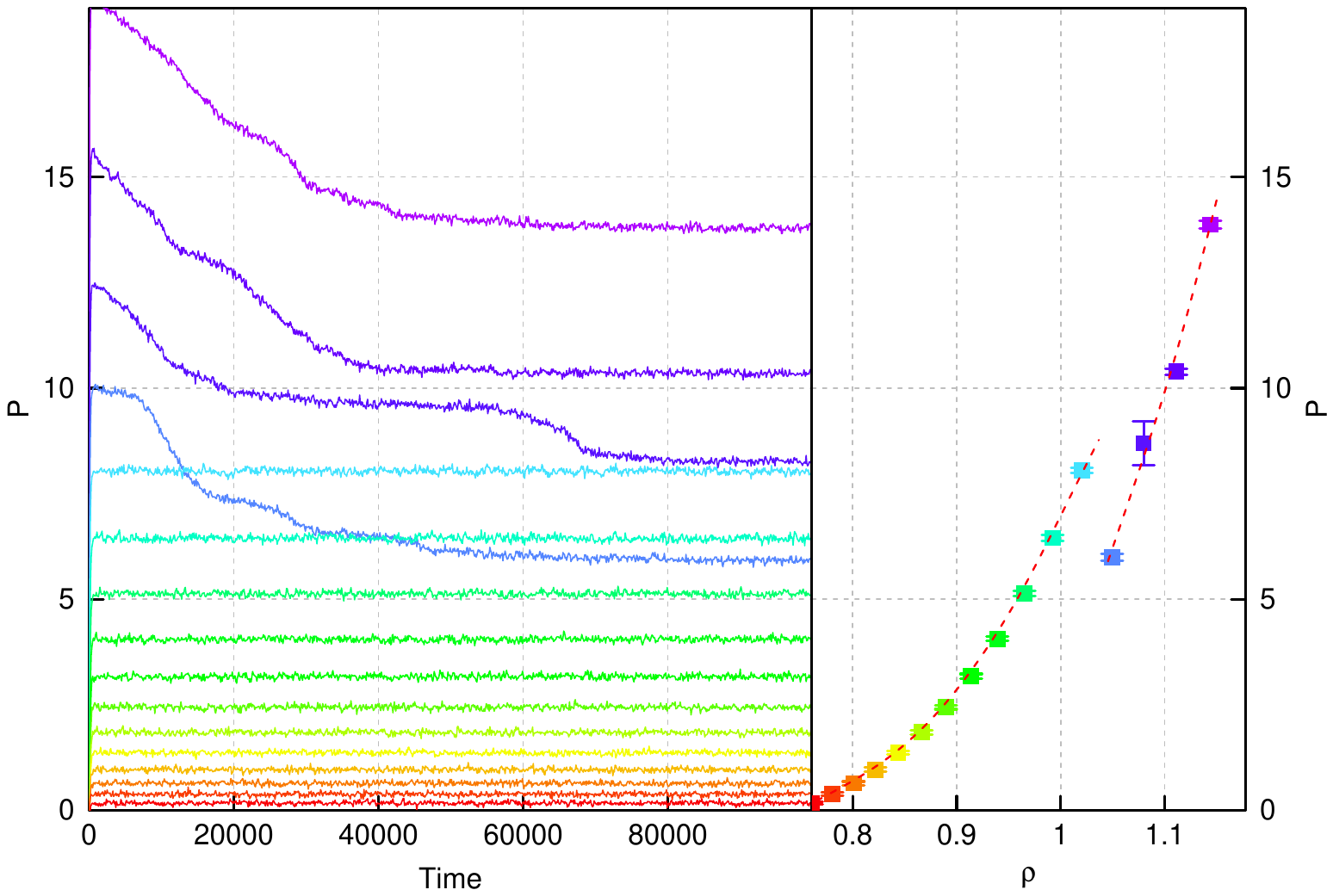}
\caption{The pressure relaxation curves (left panel) and equation of state at $T=0.9$ (right panel) in three dimensional  Lennard-Jones system. Dashed red curves at the right panel are guides to the eyes. }
\label{f:9}
\end{figure*}

\begin{figure}
\includegraphics[width=\columnwidth]{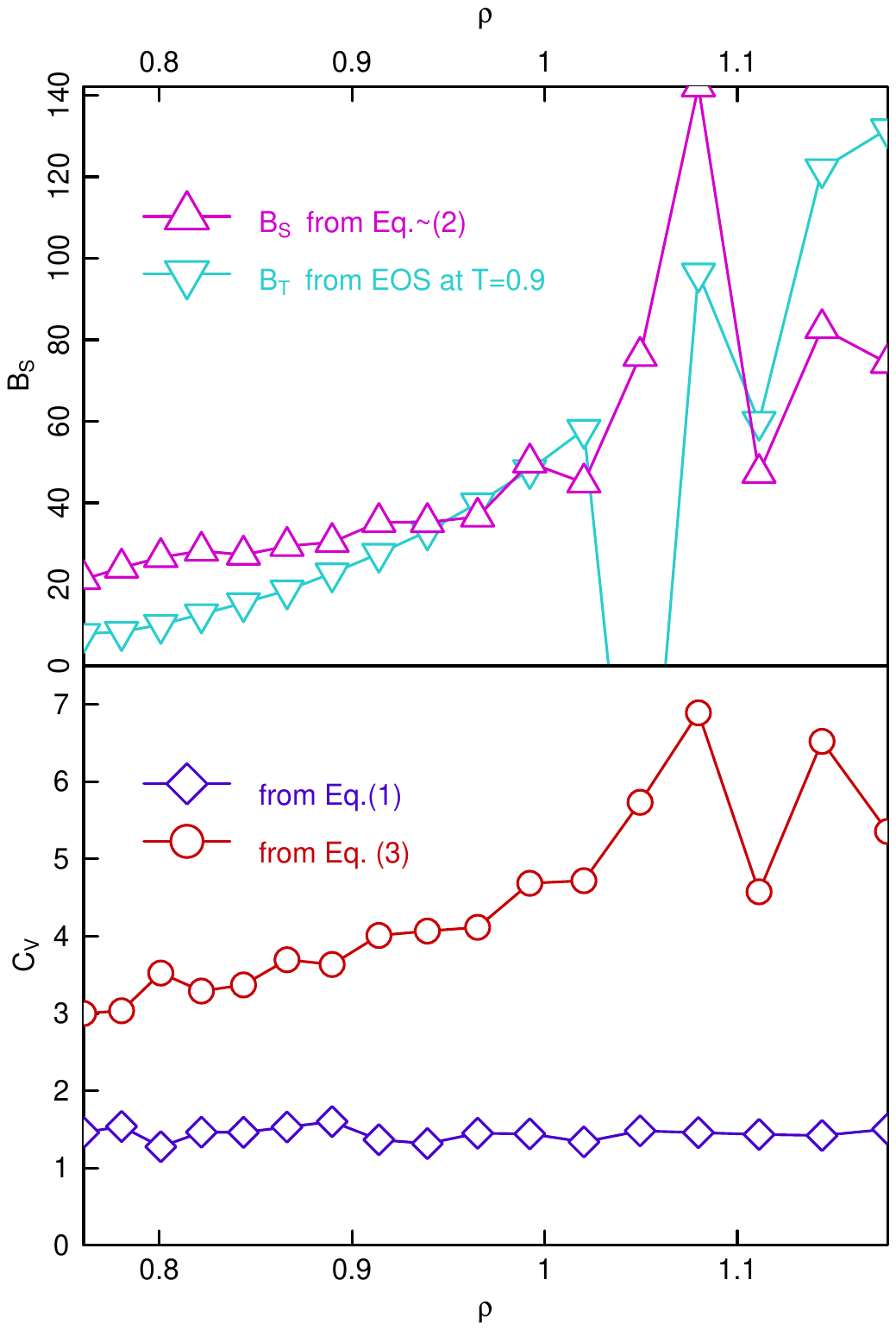}
\caption{Adiabatic bulk modulus $B_S$ and heat capacity $C_V$ evaluated from Eqs.~\ref{eq:1}-\ref{eq:3} in 3D case. Isothermal bulk modulus $B_T$ are evaluated by numerical differentiation of equation of state $P(\rho)$ along isotherm $T=0.9$.}
\label{f:10}
\end{figure}
Fluctuations of pressure and temperature of system with $N=10^4$ particles at T=0.9 are shown in Fig.~\ref{f:0} at the same set temperature $T \approx 0.9$  and different densities ($\rho=1.23$ and $\rho=0.86$) which correspond to  thermodynamic conditions deep inside the crystal phase and the liquid phase just below the transition. As it was expected, fluctuations of temperature reach about 1\% with almost normal distribution (see Fig.~\ref{f:1}). The P-$\rho$ curves in vicinity of these density-temperature conditions (near the melting transition) are shown in Fig.~\ref{f:2}. It is clear, that no van-der-Waals loops are observed in this region and with precision of pressure fluctuations they might be considered as monotonous functions increasing with the density rise. Still we need to  inspect fluctuations of temperature to make decisive conclusions.

As it is clear from distribution functions of temperature along nominal isotherm $T=0.9$ (see Fig.~\ref{f:1}) there is distinctive shift of temperatures to higher values of temperature than the one initially set at densities corresponding to solid state of the 2DLJ system. When approaching the melting transition this curve at $\rho \approx 1$ exhibits a clear crossover to lower temperatures and at these densities distribution curve is clearly bimodal. Presumably this is due to the ensemble (NVE) chosen by us where the energy of the system is kept constant but not the temperature, and slow  pressure relaxation in the crystal phase (see Fig.~\ref{f:0}) leads to deviation of temperature from the set value. Therefore temperature can not relax fast enough to follow temperature of the heat bath governed by Langevin dynamics. Still, we should stress that near melting transition the mean value of the temperature distribution function is constant and only slightly shifts above initially set value which is independent of density values. 

For comparison we provide distributions densities obtained for the same number of particles in the same conditions but calculated in the NVT ensemble (Fig.~\ref{f:3}). As expected for this sort of ensemble the distribution maximum (mean value) is closer to the initially set temperature value. However, the standard deviation is almost the same as in NVE ensemble. It is interesting to note that at $\rho \approx 1$ the distribution is also bimodal. This fact possibly catches an important feature in the dynamic of the system presumably arising from transition from the crystal to hexatic phase.

To demonstrate this idea  we provide the results for the larger NVE ensemble with $N=10^5$ particles at the same temperature $T=0.9$ in the nearest vicinity of the transition (see Fig.~\ref{f:4}). As expected, with the increase of the particles number, the values of temperature and pressure fluctuations diminish, but  slow relaxation of pressure is conserved in the crystal side of transition. However, at the upper half of time frame used for calculation of thermodynamic parameters the relaxation almost reaches its equilibrium value which is demonstrated by the density distribution of temperature fluctuations in this time region (see Fig.~\ref{f:5}). It is clear that mean temperature is only slightly above the initially set one and the  drift of the mean temperature is negligible in comparison to the  standard deviation of fluctuations. This means that the absence of the van-der-Waals loops on the pressure-density curve observed in Fig.~\ref{f:4} is not related to the drift of temperature at higher densities.

To compare our findings with previous results \cite{tsiok:jcp22} we performed calculation at higher temperature T=3.0 in NVE ensemble and $N=10^5$ particles (see Fig.~\ref{f:6}). In the work of Tsiok {\em et al.} \cite{tsiok:jcp22} calculation was done in NVT ensemble at the same $T=3.0$ temperature, but with larger particle number of  $N=256^2$. It  was stated there that the van-der-Waals loops were observed, which demonstrate first-order at least at hexatic-liquid stage of the transition. Still, the van-der-Waals loop  observed  there  had rather small amplitude (in Fig.~\ref{f:6} it looks like a plateau), being significantly smaller than the standard deviations of pressure fluctuations calculated by us. 

 By the way, fluctuations observed in the middle of transition, shown in the inset of Fig.~\ref{f:6} demonstrate convergence of pressure relaxation at longer time range ($t=2 \cdot 10^5$). From this demonstration it is evident that the time range used by us (upper half of $t=10^5$ interval) is large enough to fully converge the pressure values in the vicinity of melting transition.  

There is  qualitative difference between the two curves in  Fig.~\ref{f:6} -- plateau on the curve of Tsiok {\em et al.} \cite{tsiok:jcp22} is not only flatter than the one reported in  present work, but  also  spans a longer range of densities $\rho$. To explain   this discrepancy, we can refer to the  difference between initial  conditions in our work and Ref.~\cite{tsiok:jcp22}. In our setup initial atom coordinates  were purely random but in Ref.~\cite{tsiok:jcp22} they started from ideal triangular lattice. The latter case favors stability of crystal phase, so it persists longer below melting transitions which results in van-der-Waals loops on equation of state curve. Flatness of this curve depends on the number of system's particles,  so at the thermodynamic limit (the infinite number of particles) the van-der-Waals loop is believed to transform into ideal plateau with zero derivative in the middle.  It was demonstrated for the first time for the 2DLJ system  \cite{toxvaerd:pra81}   that  deepness of the van-der-Waals loop in the system with ``crystal'' initial state is correlated to the number of particles -- the larger the number, the flatter the curve. From this work one can see that the amplitude of van-der-Waals loop has the same order of magnitude as  system fluctuations . In Ref.~\cite{toxvaerd:pra81} for $N=256$ system both values are about 10\%, which corresponds to the rule of thumb that relative fluctuations are inversely proportional to square root of the number of particles .

There is  another argument in favor of the first-order phase transition, namely phase coexistence observed in computer experiments between hexatic and liquid phase in the region of ``van-der-Waals loop'' \cite{tsiok:jcp22}. We did not investigate this problem closely, but previous report obtained by Patashinskii {\em et al.} \cite{patash:jpcc10} in $N=2500$ 2DLJ system demonstrates that this separation has transient character, so that regions of hexatic phase and translational disordered liquid freely transform to each other. In our opinion (we agree with authors of Ref.~\cite{patash:jpcc10}) such a transformation is an indication of  continuous transition between the two phases.

There were two recent papers proposing different scenarios of the  first order phase transition   \cite{haji:pre19,li:prl20}. Rather than discussing contradictions in various simulations we would like to draw attention to the point  -- how many sigmas  the effect reported (deepness of the van-der-Waals loops) is?  In other words, we insist on the importance of being earnest with the error bars. Comparison with the noise band of pressure relaxation curve reported in our paper demonstrates that it is also much below 1-sigma. So, there is no point to discuss such a negligible effect. The effect should be as much as 3-sigma, to be statistically significant. However, effect and noise in 2DLJ system are correlated -- the larger the system size , the smaller the noise, and the shallower van der Waals loops \cite{toxvaerd:pra81}. In fact, this is discussed in   \cite{haji:pre19} as size effect-- the energetic barrier between hexatic phase and liquid diminishes as $1/\sqrt{N}$ with the system size. Hence, in the thermodynamic limit, this means that energetic barrier between two phases vanishes, that is transition is continuous, therefore it should be in full agreement with conclusions of the Berezinskii-Kosterlitz-Thouless-Halperin-Nelson-Young theory. Therefore, there is not enough evidence of a first order of hexatic-liquid transition  to overthrow existing theory of 2D melting.

Later we demonstrate that the noise observed in the 2DLJ system at large extent corresponds to the predictions of statistical physics.

\section{Discussion} 
Previously we have demonstrated the finite precision of molecular dynamic simulations. In this section we answer the second part of the question posed in the title -- what we can learn from finite precision of molecular dynamics experiments. As it follows from Eqs.~\ref{eq:1}-\ref{eq:2} from fluctuations of temperature and pressure we could estimate isochoric heat capacity and adiabatic bulk modulus.

 According to Eq.~\ref{eq:1}, we can obtain $c_V$ from standard deviation of relative fluctuation of temperature . This plot {\em vs.} density $\rho$ for temperatures in the range $T= 0.6-1.1$ is depicted in Fig.~\ref{f:7}. Overall plot is rather fuzzy, but for a fixed temperature (see dashed line in Fig.~\ref{f:7} which is a free-hand draw for $T=0.9$) we can conclude that the heat capacity is almost constant at the whole densities range.  It is  interesting to check the validity of the $N^{-1/2}$ scaling law for fluctuations. We plotted the data obtained for $N=10^5$ ensemble on the same Fig.~\ref{f:7}. It is clear that in comparison to  $N=10^4$ ensemble data there is small but regular discrepancy between them  in the range of about 15\%. It means that fluctuations in computer simulations are larger, than those predicted by statistical physics. These fluctuations could be due to other inputs,  unrelated to statistical physics, for example, like finite precision of computer calculations, discretization of time in molecular dynamics simulations, incomplete structure relaxation because of finite time of computer experiments {\em etc}. We don't want to call these  fluctuations ``unphysical'' so  we will name it  algorithmical in contrast to physical ones described by Eqs.~(\ref{eq:1}-\ref{eq:2}). It is hard to estimate the scaling of algorithmical fluctuations with the system size, but the difference between the  amplitudes of fluctuations observed in $N=10^5$ ensemble and for $N=10^4$ one (see Fig.~\ref{f:7}) suggests that its relative contribution with respect to physical fluctuations will rise with the increase of system size. May be it is not due to the absolute increase of algorithmical fluctuations but rather to decrease of physical ones. We should also mention that almost constant $c_V$ value during transition was obtained in seminal work of  Frenkel and McTague \cite{frenkel:prl79} although with higher value of $c_V=2$. This discrepancy deserves special attention. 

 Eq.~\ref{eq:1} is not new in molecular dynamics simulations but it is known in different flavor as fluctuations of internal energy $U$ (per particle):
\begin{equation}
\Delta U^2=\frac{k_B T^2 c_v}{N}
\label{eq:3}
\end{equation}
which is a consequence of Eq.~\ref{eq:1} and thermodynamic identity $\Delta U = c_v \Delta T$. For example, this formula Eq.~\ref{eq:3} was used for calculation of specific heat in the system of the  Hertzian disks in Ref.~\cite{tsiok:sm20} (the authors referenced the readers to the book \cite{frenkel02}). Generally speaking,  Eqs.~(\ref{eq:1}-\ref{eq:3}) can be regarded as thermodynamic counterpart of famous fluctuation-dissipation theorem \cite{landafshitz:v,kondrin:jcp15} where fluctuations are related to the value of certain material characteristic , such as heat capacity and adiabatic bulk modulus. However, applicability of these equations to the results obtained in computer simulations is problematic. Because of the presence of algorithmical fluctuations $c_V$ values obtained from the temperature and internal energy fluctuations (Eq.~\ref{eq:1} and Eq.~\ref{eq:3} respectively) will be different. Obviously Eq.~\ref{eq:1} will underestimate real $c_V$ and Eq.~\ref{eq:3} will overestimate it. Similarly Eq.~\ref{eq:2} should overestimate the real adiabatic bulk modulus.  Considering relative input from physical and algorithmical fluctuations to the overall noise, we should arrive to quite paradoxical conclusion, that  estimates of thermodynamic parameters from fluctuations in computer simulations will be more precise, if obtained from smaller systems, where larger physical fluctuations dominate over algorithmical ones. 

The results of estimation of specific heat and adiabatic bulk modulus from fluctuations according to Eqs.~\ref{eq:1}-\ref{eq:3} for $N=10^4$ system are depicted in Fig.~\ref{f:8}. For calculation of ``true'' specific heat we used the values of internal energy along isochores in the temperature range $T=0.6-1.1$ and numerically differentiated it by temperature. In most cases the curves require linear fit of $U$ {\em vs.} $T$ because dependencies hardly deviate from linear law. To check the validity of adiabatic bulk modulus estimation we used isothermal bulk modulus $B_T$ obtained from numerical differentiation (simple ratio of finite differences) of equation of state $P(\rho)$ taken at $T=0.9$. $B_T$ is related to adiabatic bulk modulus (which can in practice be measured by longitudinal sound velocity measurements) by equation: 
\begin{equation*}
\frac{B_S}{B_T} = 1 + \alpha_T \chi T
\end{equation*}
In solids the thermal expansion coefficient $\alpha_T$ is small and the Gr\"uneisen parameter $\chi$ is close to unity, so deviation of this ratio from unity is negligible. At the same time, this difference in liquids can amount to 20\% (due to larger $\alpha_T$) \cite{danilov:jpcb17}. Note that the adiabatic bulk modulus is always greater than isothermal. Adiabatic bulk modulus is discontinuous in the process of the first-order phase transition \cite{gromnitskaya:jetpl04}.

As it follows from Fig.~\ref{f:8}, the ``true'' $c_v$ is nicely bracketed by estimations from thermal fluctuations. At the same time ``true'' $c_v$ does not show any discontinuity at the region of phase transition but gradually diminishes from the values $c_V \approx 2$ in the solid phase (the Dulong-Petit law in 2D) to about 1.5 in the liquid phase. These calculations corroborate continuous character of the phase transition.

The case of $B_S$ is slightly more complicated. As expected $B_S$ is larger than $B_T$ in liquid and hexatic phases, but deep inside the crystal  phase there are points where estimated adiabatic bulk modulus is smaller than isothermal one (although we expected that $B_S$ should be overestimated). The only reasonable explanation for this anomaly is that there are slow physical fluctuations in the solid phase, which require longer time for their registration than our time frame. Taking into account the presence of slow pressure relaxations in solid state (see Fig.~\ref{f:4}) it is naturally to assume that slow fluctuations are also possible. To estimate the amplitude of total fluctuations we do not need to relax the system for longer times, the simple estimate would be obtained from the multiplication of the error bars of $P(\rho)$ curve in the solid phase by $\sqrt{max(B_T/B_S)} \approx 1.8$ times. Surely this operation  does not increase statistical significance of van-der-Waals loops reported in the literature earlier. However, we acknowledge that demonstration of slow fluctuations in the solid state requires careful investigation of spectral decomposition of fluctuations, that we are not ready to perform.  
 
There is one more philosophical issue. Internal energy and temperature are physical quantities, so in principle it is possible to independently measure their fluctuations in quite small systems (alike to the Brownian motion or the Nyquist noise measurement experiments). If specific heat values estimated from this measurements would differ, it would indicate the presence of algorithmical fluctuations and, therefore, it would mean positive resolution of The Matrix hypothesis \cite{matrix99}: ``Do we live in computer simulation or the world is real?'' In any case, possibility of potential experimental resolution of this question makes it  quite a valid scientific problem. 
 
\section{Comparison with 3D case}

The main purpose of this small section is brief demonstration of validity of methods used for investigation of 2D Lennard-Jones system in comparison to 3D case. There is no doubt that 3D melting  is the first-order transition. So, how would true first-order transition  look like in computer simulations? In 3D case we used the same parameters as for calculations of 2DLJ system except we had 3 dimensions. We restrict ourselves to NVE ensemble with $N=10^4$ particles. It allows us to check the convergence of finite time calculations to  the final state of transition. As in the 2D case, we start from random initial configuration that is, simulating crystallization rather than melting. It would be demonstrated that possible crystallization into polycrystalline state does not influence much the character of transition. Also by this calculation we would dispel a popular belief, that  fluctuations in 3D case are significantly smaller than in 2D case. In both cases fluctuations are governed by the same Equations \ref{eq:1}-\ref{eq:3} although with different parameters $c_V$ and $B_S$.

Equation of state at $T=0.9$ and pressure relaxation curves at different densities $\rho$ are depicted in Fig.~\ref{f:9}. The first conspicuous feature is the amplitude of van-der-Waals loop observed on the equation of state -- it is many times larger than the error bars, so the effect is statistically significant. The first order of transition also manifests itself in visual difference of pressure relaxation curves in Fig.~\ref{f:9} and that in 2D case (Fig.~\ref{f:4}). It is clear that qualitative difference of pressure relaxation curves in solid and liquid case in 3D -- the slow pressure relaxations present in solid state are  totally absent in liquid state, and gradually diminishing of slow relaxations during hexatic-liquid transition in 2D case.

Calculations of $B_S$, $C_V$ from fluctuations and $B_T$ from equation of state are depicted in Fig.~\ref{f:10}. The discrepancy between $B_S$ and $B_T$ on the one hand and $C_V$ values obtained from Eq.~\ref{eq:1} and Eq.~\ref{eq:3} respectively are similar to that observed in 2D case (Fig.~\ref{f:8}) although the final values differ.

\section{Conclusions} 

To conclude we demonstrate that the van-der-Waals loops observed on equation of state of 2DLJ system which were believed before to be an indication of first order phase transition between crystal and liquid in two dimensions are at the level of noise  and can not be reliably used for determination of transition's character. At least  this statistically insignificant effect is surely not enough to overthrow existing Nobel prize winning theories of 2D melting. The increase of the particles number  used in molecular dynamics simulation can result in attenuation of standard deviation of fluctuations, but in the same time (as it was demonstrated before \cite{toxvaerd:pra81}) it  results in flattening of  van-der-Waals loops on equation of state curves. Estimation of inherent statistical noise present in computer simulations lead us to conclusion that it is larger than predicted by the statistical physics and the difference between them (named algorithmical fluctuations) are possibly caused by the computer related issues. It was demonstrated that these fluctuations in principle could be observed in real-life physical experiments which would lead to resolution of The Matrix hypothesis \cite{matrix99}.

Data is available on reasonable request from corresponding author.
\input{2dlj-7.bbl}

\end{document}

%% file: 2dlj-7.bbl
%

%% file: 2dlj-7.bbl
\begin{thebibliography}{31}%
\makeatletter
\providecommand \@ifxundefined [1]{%
 \@ifx{#1\undefined}
}%
\providecommand \@ifnum [1]{%
 \ifnum #1\expandafter \@firstoftwo
 \else \expandafter \@secondoftwo
 \fi
}%
\providecommand \@ifx [1]{%
 \ifx #1\expandafter \@firstoftwo
 \else \expandafter \@secondoftwo
 \fi
}%
\providecommand \natexlab [1]{#1}%
\providecommand \enquote  [1]{``#1''}%
\providecommand \bibnamefont  [1]{#1}%
\providecommand \bibfnamefont [1]{#1}%
\providecommand \citenamefont [1]{#1}%
\providecommand \href@noop [0]{\@secondoftwo}%
\providecommand \href [0]{\begingroup \@sanitize@url \@href}%
\providecommand \@href[1]{\@@startlink{#1}\@@href}%
\providecommand \@@href[1]{\endgroup#1\@@endlink}%
\providecommand \@sanitize@url [0]{\catcode `\\12\catcode `\$12\catcode
  `\&12\catcode `\#12\catcode `\^12\catcode `\_12\catcode `\%12\relax}%
\providecommand \@@startlink[1]{}%
\providecommand \@@endlink[0]{}%
\providecommand \url  [0]{\begingroup\@sanitize@url \@url }%
\providecommand \@url [1]{\endgroup\@href {#1}{\urlprefix }}%
\providecommand \urlprefix  [0]{URL }%
\providecommand \Eprint [0]{\href }%
\providecommand \doibase [0]{http://dx.doi.org/}%
\providecommand \selectlanguage [0]{\@gobble}%
\providecommand \bibinfo  [0]{\@secondoftwo}%
\providecommand \bibfield  [0]{\@secondoftwo}%
\providecommand \translation [1]{[#1]}%
\providecommand \BibitemOpen [0]{}%
\providecommand \bibitemStop [0]{}%
\providecommand \bibitemNoStop [0]{.\EOS\space}%
\providecommand \EOS [0]{\spacefactor3000\relax}%
\providecommand \BibitemShut  [1]{\csname bibitem#1\endcsname}%
\let\auto@bib@innerbib\@empty
\bibitem [{\citenamefont {Hickman}\ and\ \citenamefont
  {Mishin}(2016)}]{hikman:prb16}%
  \BibitemOpen
  \bibfield  {author} {\bibinfo {author} {\bibfnamefont {J.}~\bibnamefont
  {Hickman}}\ and\ \bibinfo {author} {\bibfnamefont {Y.}~\bibnamefont
  {Mishin}},\ }\href {\doibase 10.1103/PhysRevB.94.184311} {\bibfield
  {journal} {\bibinfo  {journal} {Phys. Rev. B}\ }\textbf {\bibinfo {volume}
  {94}},\ \bibinfo {pages} {184311} (\bibinfo {year} {2016})}\BibitemShut
  {NoStop}%
\bibitem [{\citenamefont {Landau}\ \emph {et~al.}(1980)\citenamefont {Landau},
  \citenamefont {Pitaevskii},\ and\ \citenamefont {Lifshitz}}]{landafshitz:v}%
  \BibitemOpen
  \bibfield  {author} {\bibinfo {author} {\bibfnamefont {L.}~\bibnamefont
  {Landau}}, \bibinfo {author} {\bibfnamefont {L.}~\bibnamefont {Pitaevskii}},
  \ and\ \bibinfo {author} {\bibfnamefont {E.}~\bibnamefont {Lifshitz}},\
  }\href@noop {} {\emph {\bibinfo {title} {Statistical Physics}}},\ Course of
  theoretical physics\ (\bibinfo  {publisher} {Pergamon Press, Oxford},\
  \bibinfo {year} {1980})\BibitemShut {NoStop}%
\bibitem [{\citenamefont {Bystryi}\ \emph {et~al.}(2014)\citenamefont
  {Bystryi}, \citenamefont {Lavrinenko}, \citenamefont {Lankin}, \citenamefont
  {Morozov}, \citenamefont {Norman},\ and\ \citenamefont
  {Saitov}}]{saitov:ht14}%
  \BibitemOpen
  \bibfield  {author} {\bibinfo {author} {\bibfnamefont {R.~G.}\ \bibnamefont
  {Bystryi}}, \bibinfo {author} {\bibfnamefont {Y.~S.}\ \bibnamefont
  {Lavrinenko}}, \bibinfo {author} {\bibfnamefont {A.~V.}\ \bibnamefont
  {Lankin}}, \bibinfo {author} {\bibfnamefont {I.~V.}\ \bibnamefont {Morozov}},
  \bibinfo {author} {\bibfnamefont {G.~E.}\ \bibnamefont {Norman}}, \ and\
  \bibinfo {author} {\bibfnamefont {I.~M.}\ \bibnamefont {Saitov}},\ }\href
  {\doibase 10.1134/S0018151X14040063} {\bibfield  {journal} {\bibinfo
  {journal} {High Temperature}\ }\textbf {\bibinfo {volume} {52}},\ \bibinfo
  {pages} {475} (\bibinfo {year} {2014})}\BibitemShut {NoStop}%
\bibitem [{lj:()}]{lj:wiki}%
  \BibitemOpen
  \href {https://en.wikipedia.org/wiki/Lennard-Jones_potential} {\enquote
  {\bibinfo {title} {Lennard-jones potential},}\ }\BibitemShut {NoStop}%
\bibitem [{\citenamefont {Ryzhov}\ \emph {et~al.}(2023)\citenamefont {Ryzhov},
  \citenamefont {Gaiduk}, \citenamefont {Tareeva}, \citenamefont {Fomin},\ and\
  \citenamefont {Tsiok}}]{ryzhov:jetp23}%
  \BibitemOpen
  \bibfield  {author} {\bibinfo {author} {\bibfnamefont {V.~N.}\ \bibnamefont
  {Ryzhov}}, \bibinfo {author} {\bibfnamefont {E.~A.}\ \bibnamefont {Gaiduk}},
  \bibinfo {author} {\bibfnamefont {E.~E.}\ \bibnamefont {Tareeva}}, \bibinfo
  {author} {\bibfnamefont {Y.~D.}\ \bibnamefont {Fomin}}, \ and\ \bibinfo
  {author} {\bibfnamefont {E.~N.}\ \bibnamefont {Tsiok}},\ }\href {\doibase
  10.1134/S1063776123070129} {\bibfield  {journal} {\bibinfo  {journal}
  {Journal of Experimental and Theoretical Physics}\ }\textbf {\bibinfo
  {volume} {137}},\ \bibinfo {pages} {125} (\bibinfo {year}
  {2023})}\BibitemShut {NoStop}%
\bibitem [{\citenamefont {Barker}\ \emph {et~al.}(1981)\citenamefont {Barker},
  \citenamefont {Henderson},\ and\ \citenamefont {Abraham}}]{barker:pa81}%
  \BibitemOpen
  \bibfield  {author} {\bibinfo {author} {\bibfnamefont {J.}~\bibnamefont
  {Barker}}, \bibinfo {author} {\bibfnamefont {D.}~\bibnamefont {Henderson}}, \
  and\ \bibinfo {author} {\bibfnamefont {F.}~\bibnamefont {Abraham}},\ }\href
  {\doibase https://doi.org/10.1016/0378-4371(81)90222-3} {\bibfield  {journal}
  {\bibinfo  {journal} {Physica A: Statistical Mechanics and its Applications}\
  }\textbf {\bibinfo {volume} {106}},\ \bibinfo {pages} {226} (\bibinfo {year}
  {1981})}\BibitemShut {NoStop}%
\bibitem [{\citenamefont {Frenkel}\ and\ \citenamefont
  {McTague}(1979)}]{frenkel:prl79}%
  \BibitemOpen
  \bibfield  {author} {\bibinfo {author} {\bibfnamefont {D.}~\bibnamefont
  {Frenkel}}\ and\ \bibinfo {author} {\bibfnamefont {J.~P.}\ \bibnamefont
  {McTague}},\ }\href {\doibase 10.1103/PhysRevLett.42.1632} {\bibfield
  {journal} {\bibinfo  {journal} {Phys. Rev. Lett.}\ }\textbf {\bibinfo
  {volume} {42}},\ \bibinfo {pages} {1632} (\bibinfo {year}
  {1979})}\BibitemShut {NoStop}%
\bibitem [{\citenamefont {Toxvaerd}(1981)}]{toxvaerd:pra81}%
  \BibitemOpen
  \bibfield  {author} {\bibinfo {author} {\bibfnamefont {S.}~\bibnamefont
  {Toxvaerd}},\ }\href {\doibase 10.1103/PhysRevA.24.2735} {\bibfield
  {journal} {\bibinfo  {journal} {Phys. Rev. A}\ }\textbf {\bibinfo {volume}
  {24}},\ \bibinfo {pages} {2735} (\bibinfo {year} {1981})}\BibitemShut
  {NoStop}%
\bibitem [{\citenamefont {Tobochnik}\ and\ \citenamefont
  {Chester}(1982)}]{tobochnik:prb82}%
  \BibitemOpen
  \bibfield  {author} {\bibinfo {author} {\bibfnamefont {J.}~\bibnamefont
  {Tobochnik}}\ and\ \bibinfo {author} {\bibfnamefont {G.~V.}\ \bibnamefont
  {Chester}},\ }\href {\doibase 10.1103/PhysRevB.25.6778} {\bibfield  {journal}
  {\bibinfo  {journal} {Phys. Rev. B}\ }\textbf {\bibinfo {volume} {25}},\
  \bibinfo {pages} {6778} (\bibinfo {year} {1982})}\BibitemShut {NoStop}%
\bibitem [{\citenamefont {Koch}\ and\ \citenamefont
  {Abraham}(1983)}]{koch:prb83}%
  \BibitemOpen
  \bibfield  {author} {\bibinfo {author} {\bibfnamefont {S.~W.}\ \bibnamefont
  {Koch}}\ and\ \bibinfo {author} {\bibfnamefont {F.~F.}\ \bibnamefont
  {Abraham}},\ }\href {\doibase 10.1103/PhysRevB.27.2964} {\bibfield  {journal}
  {\bibinfo  {journal} {Phys. Rev. B}\ }\textbf {\bibinfo {volume} {27}},\
  \bibinfo {pages} {2964} (\bibinfo {year} {1983})}\BibitemShut {NoStop}%
\bibitem [{\citenamefont {Bakker}\ \emph {et~al.}(1984)\citenamefont {Bakker},
  \citenamefont {Bruin},\ and\ \citenamefont {Hilhorst}}]{bakker:prl84}%
  \BibitemOpen
  \bibfield  {author} {\bibinfo {author} {\bibfnamefont {A.~F.}\ \bibnamefont
  {Bakker}}, \bibinfo {author} {\bibfnamefont {C.}~\bibnamefont {Bruin}}, \
  and\ \bibinfo {author} {\bibfnamefont {H.~J.}\ \bibnamefont {Hilhorst}},\
  }\href {\doibase 10.1103/PhysRevLett.52.449} {\bibfield  {journal} {\bibinfo
  {journal} {Phys. Rev. Lett.}\ }\textbf {\bibinfo {volume} {52}},\ \bibinfo
  {pages} {449} (\bibinfo {year} {1984})}\BibitemShut {NoStop}%
\bibitem [{\citenamefont {Udink}\ and\ \citenamefont {van~der
  Elsken}(1987)}]{udink:prb87}%
  \BibitemOpen
  \bibfield  {author} {\bibinfo {author} {\bibfnamefont {C.}~\bibnamefont
  {Udink}}\ and\ \bibinfo {author} {\bibfnamefont {J.}~\bibnamefont {van~der
  Elsken}},\ }\href {\doibase 10.1103/PhysRevB.35.279} {\bibfield  {journal}
  {\bibinfo  {journal} {Phys. Rev. B}\ }\textbf {\bibinfo {volume} {35}},\
  \bibinfo {pages} {279} (\bibinfo {year} {1987})}\BibitemShut {NoStop}%
\bibitem [{\citenamefont {Chen}\ \emph {et~al.}(1995)\citenamefont {Chen},
  \citenamefont {Kaplan},\ and\ \citenamefont {Mostoller}}]{chen:prl95}%
  \BibitemOpen
  \bibfield  {author} {\bibinfo {author} {\bibfnamefont {K.}~\bibnamefont
  {Chen}}, \bibinfo {author} {\bibfnamefont {T.}~\bibnamefont {Kaplan}}, \ and\
  \bibinfo {author} {\bibfnamefont {M.}~\bibnamefont {Mostoller}},\ }\href
  {\doibase 10.1103/PhysRevLett.74.4019} {\bibfield  {journal} {\bibinfo
  {journal} {Phys. Rev. Lett.}\ }\textbf {\bibinfo {volume} {74}},\ \bibinfo
  {pages} {4019} (\bibinfo {year} {1995})}\BibitemShut {NoStop}%
\bibitem [{\citenamefont {Somer}\ \emph {et~al.}(1997)\citenamefont {Somer},
  \citenamefont {Canright}, \citenamefont {Kaplan}, \citenamefont {Chen},\ and\
  \citenamefont {Mostoller}}]{somer:prl97}%
  \BibitemOpen
  \bibfield  {author} {\bibinfo {author} {\bibfnamefont {F.~L.}\ \bibnamefont
  {Somer}}, \bibinfo {author} {\bibfnamefont {G.~S.}\ \bibnamefont {Canright}},
  \bibinfo {author} {\bibfnamefont {T.}~\bibnamefont {Kaplan}}, \bibinfo
  {author} {\bibfnamefont {K.}~\bibnamefont {Chen}}, \ and\ \bibinfo {author}
  {\bibfnamefont {M.}~\bibnamefont {Mostoller}},\ }\href {\doibase
  10.1103/PhysRevLett.79.3431} {\bibfield  {journal} {\bibinfo  {journal}
  {Phys. Rev. Lett.}\ }\textbf {\bibinfo {volume} {79}},\ \bibinfo {pages}
  {3431} (\bibinfo {year} {1997})}\BibitemShut {NoStop}%
\bibitem [{\citenamefont {Somer}\ \emph {et~al.}(1998)\citenamefont {Somer},
  \citenamefont {Canright},\ and\ \citenamefont {Kaplan}}]{somer:pre98}%
  \BibitemOpen
  \bibfield  {author} {\bibinfo {author} {\bibfnamefont {F.~L.}\ \bibnamefont
  {Somer}}, \bibinfo {author} {\bibfnamefont {G.~S.}\ \bibnamefont {Canright}},
  \ and\ \bibinfo {author} {\bibfnamefont {T.}~\bibnamefont {Kaplan}},\ }\href
  {\doibase 10.1103/PhysRevE.58.5748} {\bibfield  {journal} {\bibinfo
  {journal} {Phys. Rev. E}\ }\textbf {\bibinfo {volume} {58}},\ \bibinfo
  {pages} {5748} (\bibinfo {year} {1998})}\BibitemShut {NoStop}%
\bibitem [{\citenamefont {Wierschem}\ and\ \citenamefont
  {Manousakis}(2010)}]{wierschem:pp10}%
  \BibitemOpen
  \bibfield  {author} {\bibinfo {author} {\bibfnamefont {K.}~\bibnamefont
  {Wierschem}}\ and\ \bibinfo {author} {\bibfnamefont {E.}~\bibnamefont
  {Manousakis}},\ }\href {\doibase https://doi.org/10.1016/j.phpro.2010.01.213}
  {\bibfield  {journal} {\bibinfo  {journal} {Physics Procedia}\ }\textbf
  {\bibinfo {volume} {3}},\ \bibinfo {pages} {1515} (\bibinfo {year} {2010})},\
  \bibinfo {note} {proceedings of the 22th Workshop on Computer Simulation
  Studies in Condensed Matter Physics (CSP 2009)}\BibitemShut {NoStop}%
\bibitem [{\citenamefont {Patashinski}\ \emph {et~al.}(2010)\citenamefont
  {Patashinski}, \citenamefont {Orlik}, \citenamefont {Mitus}, \citenamefont
  {Grzybowski},\ and\ \citenamefont {Ratner}}]{patash:jpcc10}%
  \BibitemOpen
  \bibfield  {author} {\bibinfo {author} {\bibfnamefont {A.~Z.}\ \bibnamefont
  {Patashinski}}, \bibinfo {author} {\bibfnamefont {R.}~\bibnamefont {Orlik}},
  \bibinfo {author} {\bibfnamefont {A.~C.}\ \bibnamefont {Mitus}}, \bibinfo
  {author} {\bibfnamefont {B.~A.}\ \bibnamefont {Grzybowski}}, \ and\ \bibinfo
  {author} {\bibfnamefont {M.~A.}\ \bibnamefont {Ratner}},\ }\href {\doibase
  10.1021/jp1069412} {\bibfield  {journal} {\bibinfo  {journal} {The Journal of
  Physical Chemistry C}\ }\textbf {\bibinfo {volume} {114}},\ \bibinfo {pages}
  {20749} (\bibinfo {year} {2010})}\BibitemShut {NoStop}%
\bibitem [{\citenamefont {Wierschem}\ and\ \citenamefont
  {Manousakis}(2011)}]{wierschem:prb11}%
  \BibitemOpen
  \bibfield  {author} {\bibinfo {author} {\bibfnamefont {K.}~\bibnamefont
  {Wierschem}}\ and\ \bibinfo {author} {\bibfnamefont {E.}~\bibnamefont
  {Manousakis}},\ }\href {\doibase 10.1103/PhysRevB.83.214108} {\bibfield
  {journal} {\bibinfo  {journal} {Phys. Rev. B}\ }\textbf {\bibinfo {volume}
  {83}},\ \bibinfo {pages} {214108} (\bibinfo {year} {2011})}\BibitemShut
  {NoStop}%
\bibitem [{\citenamefont {Hajibabaei}\ and\ \citenamefont
  {Kim}(2019)}]{haji:pre19}%
  \BibitemOpen
  \bibfield  {author} {\bibinfo {author} {\bibfnamefont {A.}~\bibnamefont
  {Hajibabaei}}\ and\ \bibinfo {author} {\bibfnamefont {K.~S.}\ \bibnamefont
  {Kim}},\ }\href {\doibase 10.1103/PhysRevE.99.022145} {\bibfield  {journal}
  {\bibinfo  {journal} {Phys. Rev. E}\ }\textbf {\bibinfo {volume} {99}},\
  \bibinfo {pages} {022145} (\bibinfo {year} {2019})}\BibitemShut {NoStop}%
\bibitem [{\citenamefont {Li}\ and\ \citenamefont {Ciamarra}(2020)}]{li:prl20}%
  \BibitemOpen
  \bibfield  {author} {\bibinfo {author} {\bibfnamefont {Y.-W.}\ \bibnamefont
  {Li}}\ and\ \bibinfo {author} {\bibfnamefont {M.~P.}\ \bibnamefont
  {Ciamarra}},\ }\href {\doibase 10.1103/PhysRevLett.124.218002} {\bibfield
  {journal} {\bibinfo  {journal} {Phys. Rev. Lett.}\ }\textbf {\bibinfo
  {volume} {124}},\ \bibinfo {pages} {218002} (\bibinfo {year}
  {2020})}\BibitemShut {NoStop}%
\bibitem [{\citenamefont {Tsiok}\ \emph {et~al.}(2022)\citenamefont {Tsiok},
  \citenamefont {Fomin}, \citenamefont {Gaiduk}, \citenamefont {Tareyeva},
  \citenamefont {Ryzhov}, \citenamefont {Libet}, \citenamefont {Dmitryuk},
  \citenamefont {Kryuchkov},\ and\ \citenamefont {Yurchenko}}]{tsiok:jcp22}%
  \BibitemOpen
  \bibfield  {author} {\bibinfo {author} {\bibfnamefont {E.~N.}\ \bibnamefont
  {Tsiok}}, \bibinfo {author} {\bibfnamefont {Y.~D.}\ \bibnamefont {Fomin}},
  \bibinfo {author} {\bibfnamefont {E.~A.}\ \bibnamefont {Gaiduk}}, \bibinfo
  {author} {\bibfnamefont {E.~E.}\ \bibnamefont {Tareyeva}}, \bibinfo {author}
  {\bibfnamefont {V.~N.}\ \bibnamefont {Ryzhov}}, \bibinfo {author}
  {\bibfnamefont {P.~A.}\ \bibnamefont {Libet}}, \bibinfo {author}
  {\bibfnamefont {N.~A.}\ \bibnamefont {Dmitryuk}}, \bibinfo {author}
  {\bibfnamefont {N.~P.}\ \bibnamefont {Kryuchkov}}, \ and\ \bibinfo {author}
  {\bibfnamefont {S.~O.}\ \bibnamefont {Yurchenko}},\ }\href {\doibase
  10.1063/5.0075479} {\bibfield  {journal} {\bibinfo  {journal} {The Journal of
  Chemical Physics}\ }\textbf {\bibinfo {volume} {156}},\ \bibinfo {pages}
  {114703} (\bibinfo {year} {2022})}\BibitemShut {NoStop}%
\bibitem [{\citenamefont {Plimpton}(1995)}]{plimpton:jcp95}%
  \BibitemOpen
  \bibfield  {author} {\bibinfo {author} {\bibfnamefont {S.}~\bibnamefont
  {Plimpton}},\ }\href {\doibase https://doi.org/10.1006/jcph.1995.1039}
  {\bibfield  {journal} {\bibinfo  {journal} {Journal of Computational
  Physics}\ }\textbf {\bibinfo {volume} {117}},\ \bibinfo {pages} {1} (\bibinfo
  {year} {1995})}\BibitemShut {NoStop}%
\bibitem [{\citenamefont {Plimpton}\ \emph {et~al.}()\citenamefont {Plimpton},
  \citenamefont {Kohlmeyer}, \citenamefont {Thompson}, \citenamefont {Moore},\
  and\ \citenamefont {Berger}}]{lammps}%
  \BibitemOpen
  \bibfield  {author} {\bibinfo {author} {\bibfnamefont {S.}~\bibnamefont
  {Plimpton}}, \bibinfo {author} {\bibfnamefont {A.}~\bibnamefont {Kohlmeyer}},
  \bibinfo {author} {\bibfnamefont {A.}~\bibnamefont {Thompson}}, \bibinfo
  {author} {\bibfnamefont {S.}~\bibnamefont {Moore}}, \ and\ \bibinfo {author}
  {\bibfnamefont {R.}~\bibnamefont {Berger}},\ }\href {\doibase
  10.5281/zenodo.3726416} {\enquote {\bibinfo {title} {{LAMMPS Stable release
  29 September 2021}},}\ }\BibitemShut {NoStop}%
\bibitem [{cs:()}]{cs:homepage}%
  \BibitemOpen
  \href {http://www.zqex.dk/index.php/method/lammps-demo} {\enquote {\bibinfo
  {title} {Svaneborg lab computational soft-matter group: Lammps demos},}\
  }\BibitemShut {NoStop}%
\bibitem [{\citenamefont {{R Core Team}}(2012)}]{R}%
  \BibitemOpen
  \bibfield  {author} {\bibinfo {author} {\bibnamefont {{R Core Team}}},\
  }\href {http://www.R-project.org/} {\enquote {\bibinfo {title} {R: A language
  and environment for statistical computing},}\ } (\bibinfo {year} {2012}),\
  \bibinfo {note} {{ISBN} 3-900051-07-0}\BibitemShut {NoStop}%
\bibitem [{\citenamefont {Tsiok}\ \emph {et~al.}(2020)\citenamefont {Tsiok},
  \citenamefont {Gaiduk}, \citenamefont {Fomin},\ and\ \citenamefont
  {Ryzhov}}]{tsiok:sm20}%
  \BibitemOpen
  \bibfield  {author} {\bibinfo {author} {\bibfnamefont {E.~N.}\ \bibnamefont
  {Tsiok}}, \bibinfo {author} {\bibfnamefont {E.~A.}\ \bibnamefont {Gaiduk}},
  \bibinfo {author} {\bibfnamefont {Y.~D.}\ \bibnamefont {Fomin}}, \ and\
  \bibinfo {author} {\bibfnamefont {V.~N.}\ \bibnamefont {Ryzhov}},\ }\href
  {\doibase 10.1039/C9SM02262G} {\bibfield  {journal} {\bibinfo  {journal}
  {Soft Matter}\ }\textbf {\bibinfo {volume} {16}},\ \bibinfo {pages} {3962}
  (\bibinfo {year} {2020})}\BibitemShut {NoStop}%
\bibitem [{\citenamefont {Frenkel}\ and\ \citenamefont
  {Smit}(2002)}]{frenkel02}%
  \BibitemOpen
  \bibfield  {author} {\bibinfo {author} {\bibfnamefont {D.}~\bibnamefont
  {Frenkel}}\ and\ \bibinfo {author} {\bibfnamefont {B.}~\bibnamefont {Smit}},\
  }\href@noop {} {\emph {\bibinfo {title} {Understanding molecular simulation
  (From Algorithms to Applications), 2nd Edition}}}\ (\bibinfo  {publisher}
  {Academic Press},\ \bibinfo {year} {2002})\BibitemShut {NoStop}%
\bibitem [{\citenamefont {Kondrin}\ \emph {et~al.}(2015)\citenamefont
  {Kondrin}, \citenamefont {Brazhkin},\ and\ \citenamefont
  {Lebed}}]{kondrin:jcp15}%
  \BibitemOpen
  \bibfield  {author} {\bibinfo {author} {\bibfnamefont {M.~V.}\ \bibnamefont
  {Kondrin}}, \bibinfo {author} {\bibfnamefont {V.~V.}\ \bibnamefont
  {Brazhkin}}, \ and\ \bibinfo {author} {\bibfnamefont {Y.~B.}\ \bibnamefont
  {Lebed}},\ }\href {\doibase 10.1063/1.4914185} {\bibfield  {journal}
  {\bibinfo  {journal} {The Journal of Chemical Physics}\ }\textbf {\bibinfo
  {volume} {142}},\ \bibinfo {pages} {104505} (\bibinfo {year}
  {2015})}\BibitemShut {NoStop}%
\bibitem [{\citenamefont {Danilov}\ \emph {et~al.}(2017)\citenamefont
  {Danilov}, \citenamefont {Pronin}, \citenamefont {Gromnitskaya},
  \citenamefont {Kondrin}, \citenamefont {Lyapin},\ and\ \citenamefont
  {Brazhkin}}]{danilov:jpcb17}%
  \BibitemOpen
  \bibfield  {author} {\bibinfo {author} {\bibfnamefont {I.}~\bibnamefont
  {Danilov}}, \bibinfo {author} {\bibfnamefont {A.}~\bibnamefont {Pronin}},
  \bibinfo {author} {\bibfnamefont {E.}~\bibnamefont {Gromnitskaya}}, \bibinfo
  {author} {\bibfnamefont {M.}~\bibnamefont {Kondrin}}, \bibinfo {author}
  {\bibfnamefont {A.}~\bibnamefont {Lyapin}}, \ and\ \bibinfo {author}
  {\bibfnamefont {V.}~\bibnamefont {Brazhkin}},\ }\href {\doibase
  10.1021/acs.jpcb.7b05335} {\bibfield  {journal} {\bibinfo  {journal} {The
  Journal of Physical Chemistry B}\ }\textbf {\bibinfo {volume} {121}},\
  \bibinfo {pages} {8203} (\bibinfo {year} {2017})}\BibitemShut {NoStop}%
\bibitem [{\citenamefont {Gromnitskaya}\ \emph {et~al.}(2004)\citenamefont
  {Gromnitskaya}, \citenamefont {Stal'gorova}, \citenamefont {Yagafarov},
  \citenamefont {Brazhkin}, \citenamefont {Lyapin},\ and\ \citenamefont
  {Popova}}]{gromnitskaya:jetpl04}%
  \BibitemOpen
  \bibfield  {author} {\bibinfo {author} {\bibfnamefont {E.}~\bibnamefont
  {Gromnitskaya}}, \bibinfo {author} {\bibfnamefont {O.}~\bibnamefont
  {Stal'gorova}}, \bibinfo {author} {\bibfnamefont {O.}~\bibnamefont
  {Yagafarov}}, \bibinfo {author} {\bibfnamefont {V.}~\bibnamefont {Brazhkin}},
  \bibinfo {author} {\bibfnamefont {A.}~\bibnamefont {Lyapin}}, \ and\ \bibinfo
  {author} {\bibfnamefont {S.}~\bibnamefont {Popova}},\ }\href {\doibase
  10.1134/1.1851642} {\bibfield  {journal} {\bibinfo  {journal} {JETP Letters}\
  }\textbf {\bibinfo {volume} {80}},\ \bibinfo {pages} {597} (\bibinfo {year}
  {2004})}\BibitemShut {NoStop}%
\bibitem [{\citenamefont {Wachowski}\ and\ \citenamefont
  {Wachowski}(1999)}]{matrix99}%
  \BibitemOpen
  \bibfield  {author} {\bibinfo {author} {\bibfnamefont {L.}~\bibnamefont
  {Wachowski}}\ and\ \bibinfo {author} {\bibfnamefont {A.}~\bibnamefont
  {Wachowski}},\ }\href@noop {} {\enquote {\bibinfo {title} {{The Matrix
  (film)}},}\ } (\bibinfo {year} {1999})\BibitemShut {NoStop}%
\end{thebibliography}
